# Symmetrical laws of structure of helicoidally-like biopolymers in the framework of algebraic topology. IV. Topological stability of alpha helix and DNA structures.

M.I.Samoylovich[1], A.L.Talis[2]

[1]Central scientific research institute of technology "Technomash", Moscow
E-mail: samoylovich@technomash.ru
[2]Institute of Organoelement Compounds of Russian Academy of Sciences, Moscow

**ABSTRACT**

The structural parameters of $\alpha$-helix and some forms of DNA-structures are determined by methods of algebraic topology. These structures are locally periodic and correspond to the bifurcation point for minimal surfaces given by Weierstrass representation. The index of an unstable surface equals zero for them. As predicted by a theory of catastrophes, formation of such structures corresponds to lifting of configurational degeneration. Formation turns by half- turns from different helices in the implementation of double-structure as a single system, may indicate a kind of local crossing-over between motherboard and paternal helices (at least for some part of the helix). The role of the exchange have yet to understand given the impossibility of exchange between different turns spirals going counter rotating and shifted axis.

## 1. Introduction

Living nature avoids crystalline ordering type in the 3D Euclidean space $E^3$ by making at transition from invariance with respect to an infinite translational lattice to local periodicity, determined by invariance with respect to constructions of algebraic topology [1], which has been realized in special (helicoid-like) linear constructions. Notable discovery double helix structure of DNA has answered the question about how such systems are built. This has also allowed to approach the answer to the question why such systems are extraordinarily stable and how their structural features are related to geometric properties of our space and chemical properties of their constituent molecules (atoms), which is essentially the subject of the present work.

A distinctive feature of such biological systems is that for the one has to consider molecule packings rather than atomically-generated lattices. A basic difference between atomically-generated lattices and packings is that for the former their electron subsystem is described by a Brillouin zone (Voronoi polyhedron of the reciprocal lattice), and for the latter the electron subsystem of each element of the packing is described not by the Brillouin zone of the lattice, but rather by interior structure of a packing element. In descriptions of such systems it becomes harder to use of minimal surfaces(with zero mean curvature), that also are critical points of area of the surface which bounds certain volume, but not necessarily are stable and not necessarily are critical points of functionals such as volume.

Notable discovery by J. Watson [2] concerning double helix structure of DNA has answered the question about how such systems are built. This has also allowed to approach the answer to the question why such systems are extraordinarily stable and how their structural features are related to geometric properties of our space and chemical properties of their constituent molecules (atoms), which is essentially the subject of the present work. All non-congruent complete ruled (this factor is essential when using polyhedral constructions) surfaces can be realized as a one-parameter helicoids with pitch as a parameter [3]. A catenoid determined by one of such surfaces is locally isomorphic to a helicoid (Fig. 1.) that is used in the discussion of the bifurcation.

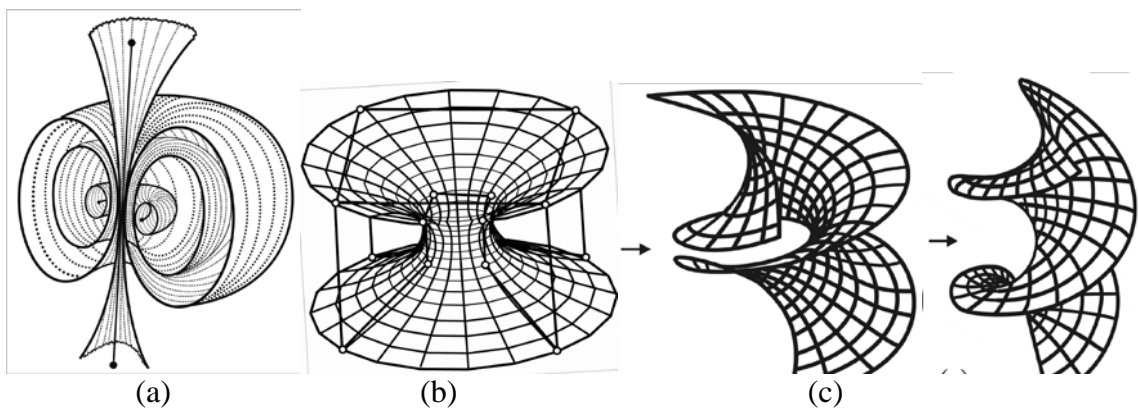

(a) (b) (c)

Figure 1
*(a) A family of tori which are coaxial with the spherical torus. This family fills up the volume of the 3-dimensional sphere $S^3$ (Figure 8 in [10]).*
*(b) The projection of {4, 3, 3} polytope vertices on the catenoid (Figure 9b in [10]), which is determined by the family of tori (Figure 1.a).*
*(c) The transformation of catenoid (b) into helicoid via an intermediate surface (wound over the catenoid) – joining of catenoid and helicoid surfaces (Figure 26 in [3]).*

In order to describe such systems (manifolds) one uses such concepts as critical (degenerate or non-degenerate), topologically regular or topologically irregular (bifurcation) points. In order to describe complex systems it is necessary to introduce multidimensional spaces (constructions); the latter implies that finally turning to description of 3D structures, one unavoidably encounters manifold maps of the form f: M→N. At the same time, the problem is to maximally preserve topological properties of the constructions in question, for which, in particular, a Morse transformation is used, when on a manifold under study a Morse function is given as a function which determines all critical points of the manifold (M).
In this paper we show that formation and stability of such structures corresponds to lifting configurational degeneration and existence of a bifurcation point.

## 2. Features of the structure of $\alpha$-helix

The secondary structure of a protein is largely determined by stiff covalent bonds in a polypeptide chain, as well as by hydrogen bonds between its side groups. An important role is played also by steric interactions of molecules, related to their sizes and shapes, imposing rigid structural limitations on positions of molecules in space. Proteins can also be considered a dense packing of more or less spherical units – amino acids, approximated by a packing of tetrahedra. The densest packing of regular tetrahedra is achieved in a 4D polyhedron (polytope) {3, 3, 5}, whose substructures are related to the α-helix as shown in [4]. As it is known, the α-helix contains a realization of the crystallographic axis 18/5 proposed by Pauling [5] with the helical rotation angle equal to $100°=360°/ (18/5)$. At the same time, hydrogen bonds that stabilize the $\alpha$-helix [6-7] appear between the *i*- amide and *(i+4)*-th carbonyl groups of residues. However, even 60 years after the publication of Pauling's paper, the symmetry-based justification is still needed for the fact that there are 3.6 amino-acid residues per turn in the $\alpha$-helix.

We have proposed a partial solution to this problem based on the derivation of special (on geometrical and topological properties) class structures in algebraic geometry. These structures defined by a set of topological properties of systems such as complete minimal surface of revolution type catenoid (the only non-planar surface with zero mean curvature) and the helix, as well as our space, taking into account the structural and geometric properties of molecules (atoms), forming such structures.

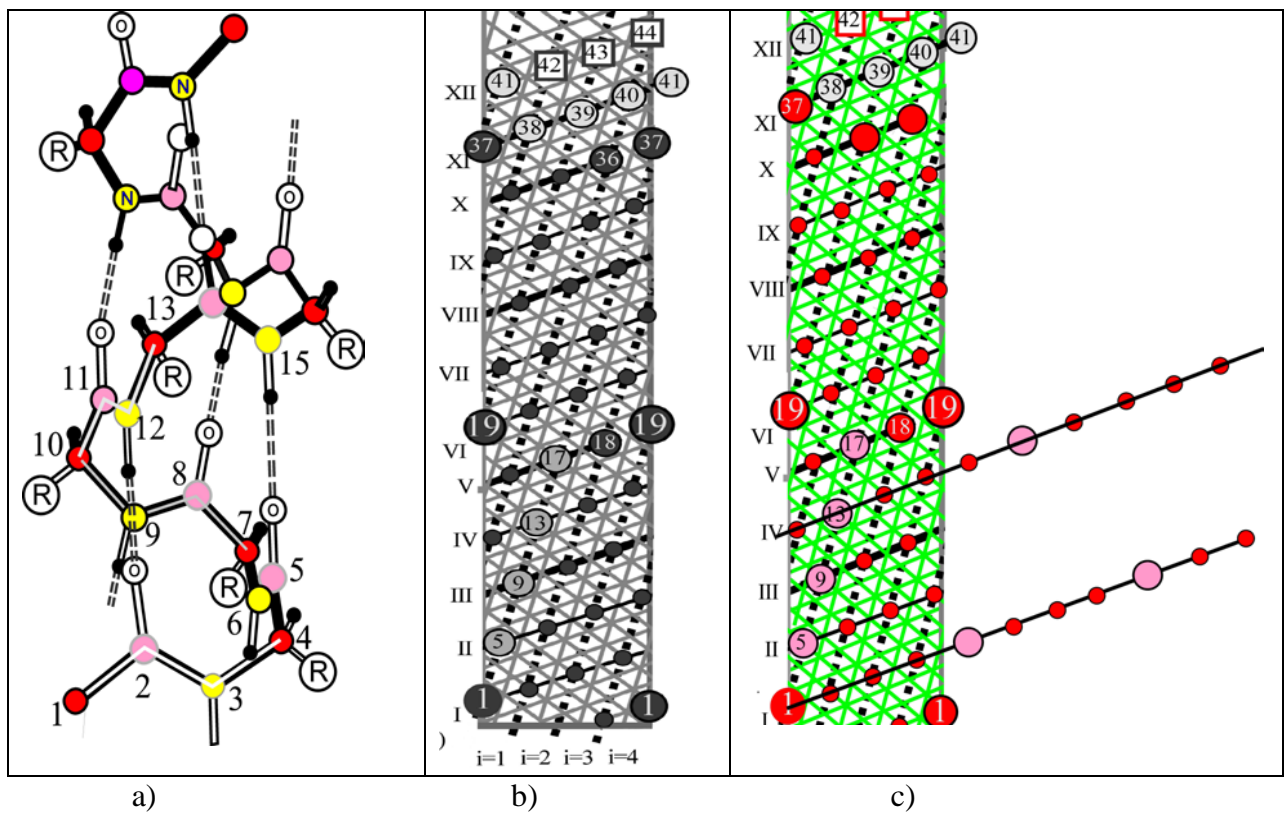

Figure 2
*(a) The structure of α-helix, pendant groups are designated by letter R, hydrogen bond shown by dotted line (Fig.7-5 in [7]). Red, pink, yellow, white and black balls represent respectively $C_\alpha$, $C'$, N, O and H atoms.*
*(b) The development of a locally cylindrical approximation of the α-helix having the 36/10 axis shown in (a). Forty-four $C_\alpha$ atoms are disposed on 12 turns designated by Roman numerals. Atoms with equal numbers are identified. Each turn designated by the thick line contains three atoms, each of the rest turns contain four atoms. Atoms i and i+4 belong to one of the four dotted straight i-lines, i=1, 2, 3, 4 (taking in the account for identifying the vertical edges of stripe).*
*(c) The development of a locally cylindrical approximation of tripled α-helix having with 11 and 10 vertices on turns.*

The conformation of the α-helix (Fig.2.a) is stable, therefore may be used its cylindrical approximation, in which the centers of congruent elements of the packing coincide with the $C_\alpha$ atomic positions or conditional molecule centers. Generally accepted data concerning the structure of the α-helix [7-8] allow one to assume, that *L* residues are equally distributed both over p turns of the "main" helix, as well as over 4 helices, which we will call *i*-helices, *i*=1,2,3,4. Each of the *i*-helices corresponds to a linear substructure of hydrogen bonds, hence a cylindrical approximation of an α-helix (Figs.2b, 3a) may be considered a result of multiplication by a screw axis L/p of a starting *i*-helix of L/4 residues.

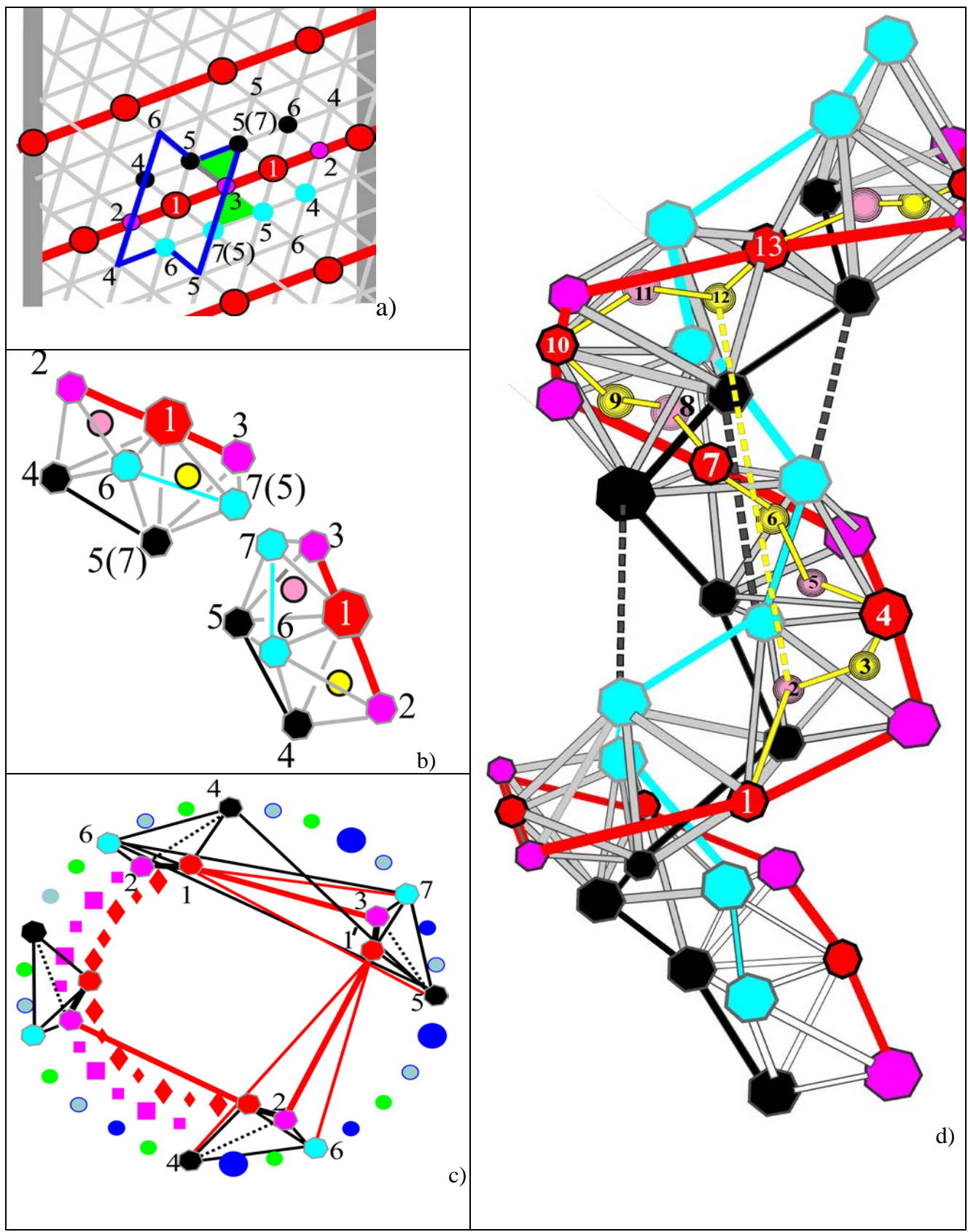

Figure 3

*(a) The development of a locally cylindrical approximation of the α-helix having the 40/11 axis as the joining of maps {3', 6}$^6_{2, 1}$. The triangles 3-5-7 is common to two neighbouring maps.*

*(b) Identifying the vertices with equal numbers of map {3, 6}$^6_{2,1}$ in Figure 3a determines a 7-vertex face-to-face joining of four regular tetrahedra with the common vertex 1. Two such joining of four* *regular tetrahedra may be united (as manifolds) by their common face 3-5-7. Shaded spheres placed in the interior of 1-2-4-6 and 1-3-5-7 tetrahedra.*

*(c) Between the tetrahedra 1-2-4-6 and 1'-3-5-7 is face-to-face joining of three regular tetrahedra with a common vertex 1. The tetrahedron 1-2-4-6 and these tetrahedra 1-4-6-5, 1-6-5-7 u 1-5-7-3 form 7-vertex union $\{3, 6\}^6_{2,1}$ of four regular tetrahedra.*

*(d) The helix packing of 7-vertex joining of tetrahedra (collected by their faces of the type 3-5-7), which is determined by identifying equal vertices of the flat development in Figure 3a. White edges delineate the 7-vertex joining of four tetrahedra which is shown in the bottom part in Figure 3b. The red edges join the red centres of the 7-vertex joining of 4 tetrahedra and pink vertices. They form a spiral, determined by the 40/11 axis. The black and blue vertices form two others 40/11 spirals. The vertices included into the $4_{13}$ cycle of the α-helix (vertices of type 1 and shaded spheres in Figure 3b), are numbered in accordance to (3) and Figures 3b, 3c. The dot-and-dash line, which is closing the $4_{13}$ cycle, connects vertices 2 and 12. Atoms O and H disposed on this dot-and-dash line and realizing the hydrogen bond are not shown. The dot-and-dash line is parallel to the dotted line connecting the nearest white and light-grey vertices.*

In contrast to 36/10, the axis 40/11 is expressed by a periodic decimal fraction and corresponds to the symmetry of the polytope {q(2·24)}. The orbit of the axis 40/11 is the union of the 4 orbits of the axis 10/1 (it is approximated by an axis close to the screw axis $10_1$), which is also a relation between axes of the cylindrical approximation of the α-helix. The helix 40/11, ensuring the existence of the four i-helices, is followed by the helix 44/12 = (40+4)/(11+1), which is the quadruple of the helix 11/3 (Fig.2.b). It is exactly 11/3 that corresponds to the average length of observed α-helices in globular proteins [6]. The one closest to 44/12 is the helix 45/12, which is a triple helix 15/4 with rotation by $96°$ (Fig.4b).

For observed lengths of α-helices in globular proteins relative maxima are selected corresponding to 7, 11 and 15 residues on two, there and four complete turns [9-14]. They can be viewed as a result of separation of 40=7+7+11+15 vertices of the helix 40/11=3.63(63), into cycles (sub-complexes, situated on 2, 2, 3 and 4 turns. The substructures, put into correspondence with such cycles, may be characterized by axes 7/2, 7/2, 11/3 and 15/4, and the properties of an averaged axis common to them as: (7/2 + 7/2+ 11/3 + 15/4)/4 =3.60416(6), which gives the experimentally determined [5] rotation by $100°$ for an α-helix by the axis 18/5.

Helix of simplicial complexes may be obtained upon transition (in a cover over a bouquet Fig.4c) from the base $S^1$ to a discrete system of points on the helix 40/11, and from the fiber $S_j^2$ to the simplicial complex $\sigma_j$:

$$S^1 \to <40/11|11h/40> v_0 \leftrightarrow \{V_1, V_2\}_\alpha,$$
$$S_j^2 \to \sigma_j \to PG(2,2) \to \{3', 6\}^6_{2,1} \leftarrow \{e_1, e_2\} \qquad (1),$$

where $<40/11|11h/40>$ is a helical rotation by $99° =11·360°/40$ with a shift along the axis by 11h/40. Upon mapping (1) the base point $j$, where the complex $\sigma_j$ is "attached", maps into the node $V_j$ of the strip $\{V_1, V_2\}_\alpha$, which is also the centre of a flat development of the map $\{3', 6\}^6_{2,1}$ of 10 isosceles triangles (Figs. 3a, 3b). Correspondingly, a cylindrical approximation of the α – helix with pitch h=5, 4 Å and radius r=2, 25 Å is determined by the relation:

$$<40/11|11h/40>^i (\{3', 6\}^6_{2,1})_1 \cong (\{3', 6\}^6_{2,1})_i \qquad (2),$$

where the center of the map $(\{3', 6\}^6_{2,1})_j$ is a node $V_j$ of the strip $\{V_1, V_2\}_\alpha$, which coincides with the $j^{th}$ atom $C_\alpha$. The triangle $\{3'\}$ is formed by the vectors $e_1$ and $e_2$ with the angle $\chi \approx 55°$ between them.

The map $(\{3', 6\}^6_{2,1})_j$ is a flat development of the union by faces of 4 tetrahedra sharing a vertex $V_j$ (Fig. 3.a b), which represent the simplicial complex $\sigma_j$ from the relation (1). This simplicial complex is a special one because it is embedded (under inessential deformations) both into an icosahedron, as well as into the Coxeter - Bordijk helicoid of regular tetrahedra [4]. Thus, the relation (1) determines a flat development of a packing of simplicial complexes, satisfying the requirement of topological stability (1) as well as experimental data concerning the structure of the α - helix [6-7] .In other words, joining of simplicial complexes (1) determines conditional centres of molecule in α-helix packing.

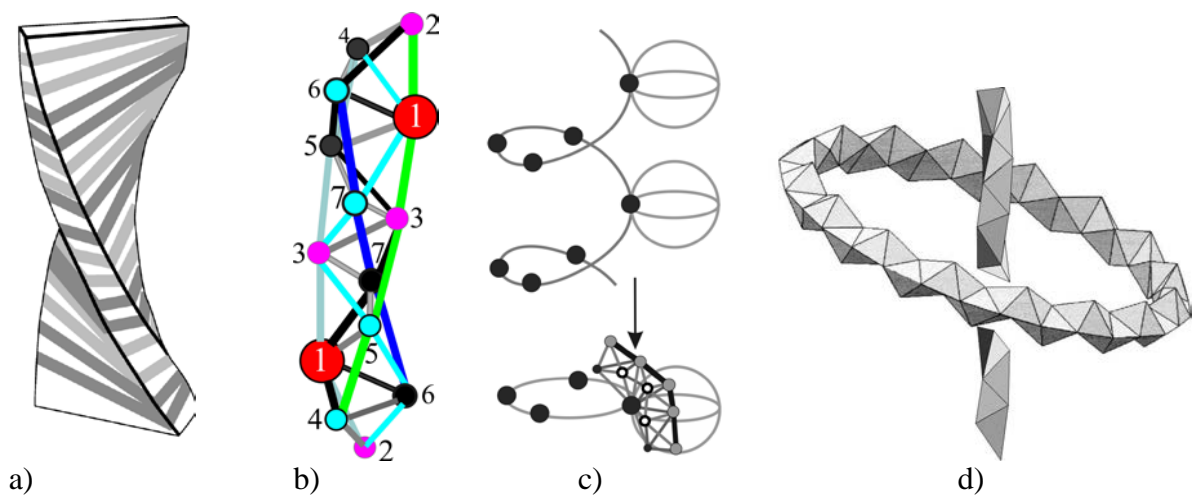

a)          b)          c)          d)

Figure 4
*(a) Joining of rods by helicoidal law [32].*
*(b) Green, dark-blue, blue-green; blue, black; grey spirals represent respectively 10/1, 15/4, 30/11 axes in the Boerdijk–Coxeter helix. Two 7-vertex simplicial complexes (with centers at large red circles) are joined by a connected sum – the three tetrahedra between them.*
*(c) Representation of a cover over a bouquet of the circle $S^1$ and the sphere $S^2$ in the form of spheres attached to a screw line. On every sphere shown is a solid common point with the screw line, corresponding to a point of a manifold on $S^1$. (Figure 103b in [14]).This point is common vertex in joining of two 7-vertex complexes (b).*
*(d) The toroidal joining of 96 tetrahedra [31], the Boerdijk–Coxeter helix (b) is located on the center of this torus.*

Under projection of 120 vertices of {3, 3, 5} onto the plane these vertices coincide with the vertices of 4 concentric 30-vertex polygons A, B, C, D. Vertices of polygons A and D coincide with the vertices of projections "symmetric" Petrie polygons of {3,3,5}. "Asymmetric" Petri polygons have the vertices of type A, B, C and B, C, D [12]. Three consecutive edges of such "asymmetric" Petri polygon connect the vertices of a regular tetrahedron in {3, 3, 5}.For instance, there are vertices B, C and two vertices A of the starting tetrahedron, where vertices A belong to different helices 15/4 (Figs.3c, 4b). Axis 15/4 translates starting tetrahedron 1, 2, 4, 6 in congruent to it tetrahedron 1', 3, 5, 7. Between the tetrahedra 1-2-4-6 and 1′-3-5-7 is face-to-face joining of three regular tetrahedra with a common vertex 1. The tetrahedron 1-2-4-6 and these three tetrahedra 1-4-6-5, 1-6-5-7, 1-5-7-3 form 7-vertex joining of four regular tetrahedra (simplicial complex) - the intersection of the icosahedron (vertex figure of the polytope {3, 3, 5}) with Petri polygon of {3, 3, 5}. The development of this 7-vertex joining of tetrahedra corresponds to irregular map $\{3, 6\}^6{}_{2,1}$ (Figure 3a,b,c), which embedded in regular map $\{3,6\}_{2,1}$. The map $\{3, 6\}_{2,1}$ is dual to map $\{6,3\}_{2,1}$ - incidence graph of minimal finite projective geometry PG (2,2).

A face-to-face joining of two 7-vertex simplicial complexes is the 11– vertex complex, which corresponds to three α - helix turns. It may be put into correspondence with the helicoid 11/3 (Fig. 2b) and has a point common for $S^1$ and $S^2$ in a cover over a bouquet (Fig 4c). Discarding four points common for $S^1$ and $S^2$ while uniting according to the law (1) four such 11-vertex complexes allows one to obtain 40 points, corresponding to the orbit 40/11. These 40 points assigned to the {160} polytope. Polytope {160} is given on the $S^3$ as cover over a bouquet, while the cover itself is the structural elements of this polytope.

Situated in a vertex of an α-helix and common to the union of 4 tetrahedra, the atom $C_\alpha$, is 4-coordinated, which determines positioning of the atoms $N$ and $C'$ inside the "exterior" tetrahedra of the union, to the left and to the right of $C_\alpha$ (Fig. 7-5 in [7]). Such decoration of the simplicial complex leads to formation of the i- link (N - $C_\alpha$ - C')$_i$ of a polypeptide chain, and ensures the assembly of the α helix:

$$-(N- [C_\alpha - C')_i–(N- C_\alpha - C')_{i+1} –(N- C_\alpha - C')_{i+2}– (N- C_\alpha - C')_{i+3}–(N -C_\alpha] - C')_{i+4} – \quad (3),$$

in which $C_\alpha$ from the *i*th complex is related to $C_\alpha$ from (i+4)-th complex by the transformation $10_1 \rightarrow (40/11)^4$. The cycle forming between the mentioned $C_\alpha$ contains 13 atoms and is selected by square brackets. Jumping ahead, let us note that replacing in such a cycle the first $C_\alpha$ for $O$, and the last for $H$, we obtain the cycle $4_{13}$, characterizing an α - helix. The number 13 (in the definition of cycle by [7], in fact, gives the number of atoms in a cycle, but 4 must be determined as the degree of the axis 40/11, mapping the i-th $C_\alpha$ into the (i+4) th $C_\alpha$.

The mapping of a polypeptide chain (3) by a flat development of a packing of tetrahedra presents a chain of isosceles triangles with common vertices, with $C_\alpha$, corresponding to these common vertices (Fig. 3a). At the same time, the atoms $C'$ and $N$ are positioned on the midpoints of the bases of triangles (or in other positions selected by symmetry in triangles). On each of the lines joining $C'_i$ and $N_{i+4}$, i=1, 2… there are 2 positions, special by symmetry, of the lattice $\{e_1, e_2\}$, which correspond to positions of the atoms $O_i$ and $H_{i+4}$. Replacing $(C_\alpha)_i$ and $(C_\alpha)_{i+4}$ for $O_i$ and $H_{i+4}$, we get the cycle $4_{13}$ characterizing the α-helix – a sequence of 13 vertices, numbered in fig.7-3. [7]. The definition [7] of the cycle $4_{13}$ characterizing the α - helix is incomplete, because it does not map the α - helix as an orbit of an axis of helical rotation 40/11. According to (1), (2), the shortest distance between the atoms $C_\alpha$ is determined by the transformation $<40/11>^1$, and the second distance between the atoms $C_\alpha$ by the transformation $<40/11>^4 \rightarrow 10/1$. Thus, the given cycle, as a closed piece of the chain (3), must be determined by the parametric axis 40/11:

$$4_{13}=\{\bigcup_{i=0}^{4}< 40/11 >^i (N, C_\alpha, C')_1 \mid N_1=C_5'=0, (C_\alpha)_1 \rightarrow O_1 \cup H_5 \leftarrow (C_\alpha)_5=<10/1>(C_\alpha)_1\} \quad (4).$$

In (4) the number 4 in the notation of the cycle $4_{13}$ is defined as the power of the axis 40/11, which maps the i-th link (N, $C_\alpha$, C′) into the (i+4)-th one and selects a union of 5 links (N, $C_\alpha$, C′)$_i$ ,i=0,1,2,3,4,5. Since 10/1 connects the initial and the final atoms $C_\alpha$, out of the five links (N, $C_\alpha$, C′) the initial $N_1$ and the final $C_5'$ atoms must be discarded. The carbon atoms $(C_\alpha)_1$ and $(C_\alpha)_5$ linked to $C_1'$ and $N_5$ are tetra-coordinated, a physical bond between them is impossible, hence the atoms $O_1$ и $H_5$ (also connected with $C_1'$ and $N_5$) enter the cycle instead of them. A bond between the atoms $O_1$ and $H_5$ is possible, and it closes the cycle of 13 atoms, which characterizes the α - helix (Fig. 3d).

Within our approach the α-helix corresponds to a substructure of a polytope {q (2·24)}, which is mapped into an octagonal face of the truncated cuboctahedron. Thus, the union of its 3 closest octagons (Fig. 5a) is in correspondence with a super-helix formed by the α-helices, whose symmetry is determined by the symmetry of a polytope. A scheme of the super-spiral is shown in Fig.5.b (Fig.11-3. in [7]). Within the triple of α-helices, which is characterized by the axes 40/11 and corresponds to the octagons in Fig. 5a, there appears a channel, which is characterized (parametrically) by the axis 30/11 and corresponding to the hexagon in Fig.5a. At the same time, between pairs of channels 40/11 there appear channels 40/9, which are in correspondence with squares (Fig.5a). Analogous relations within the approach being developed may also be obtained for other super-helices

The bifurcation point for α-helix was discussed in [19, 20]. Below we describe realization of DNA structure as a chain of lifting configurational degeneracies.

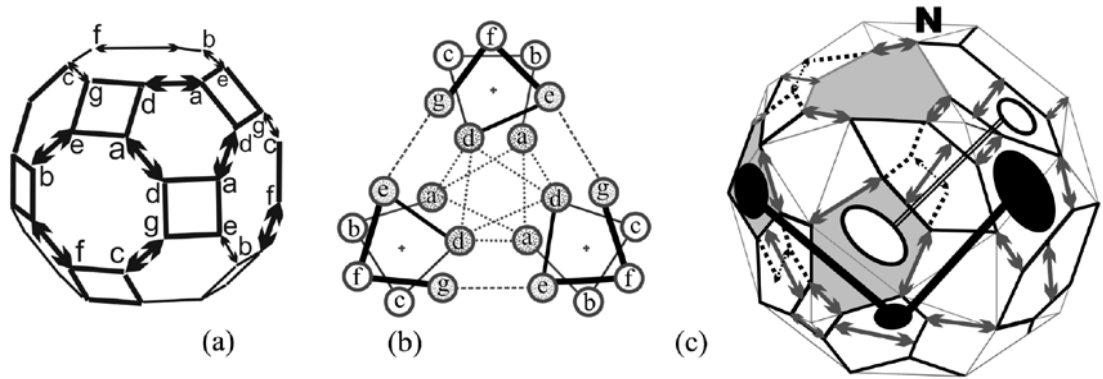

*Figure 5*
*(a) A truncated cubooctahedron with letters marking out seven vertices of the nearest to each other octagonal faces.*
*(b) The scheme of super-helix composed from three α-helices (Figure11-3 in [7]) shown as a-b-...-g helices and corresponding to octagonal faces in (a). The channels appear in the interior of three helices and between α-helices pairs are corresponding to hexagonal and square faces in (a).*
*(c) The polyhedron with $2^2 \cdot 24$ vertices and 12 heptagons, 24 pentagons, 8+6 hexagons (of two types) as faces. Three grey heptagons around a hexagon appear by transforming three octagons in (a). This transformation is effected by $90°$ rotations of dotted arrowed edges. Black and white ovals single out a parts of zigzag chains composed from three pairs of pentagons and heptagons. The «north» pole of the polyhedron designated by the letter N.*

### 3. Features of the structure of DNA

The axis 40/11 maps (parametrically) the triples of atoms $C_\alpha$, $C'$ *and* $N$ into each other; at the same time, the atoms listed do not map into each other. Let us assume that all atoms $C'$ and $N$ are projected onto the helix containing the centers of the atoms $C_\alpha$, and the centers of any two adjacent atoms are mapped into each other by one and the same (up to adjointness) transformation. The mappings of the centers (projected onto a helix) of the atoms $C_\alpha$ and $C'$, $C'$ and $N$, $N$ and $C_\alpha$ are in correspondence with three non-unit involutions in the Chevalley group of type $G_2$. Correspondingly, such a homogeneous helix is going to be mapped onto itself by the integral axis $120/11 \to (40 \cdot 3)/11$, performing the rotation by $33°$. The non-integer axis 120/11 practically coincides with a screw (non-crystallographic) axis 11=121/11, with 11 transitive elements per turn. For an α-helix the ratio $h/r \approx 2.4$ was determined for the bifurcation point of the catenoid, which (after a series of steps) allows one to (locally) define elements of the helicoid [19,20], and, consequently, given by the ends of its generatrix (norms) the double helix (Fig 4a).

Upon selecting on a sphere "the north and the south" disks (around the north and the south hexagons) and subsequently gluing the disks, two zigzag-like chains of 3 pairs of elements (heptagons and pentagons) may be put into correspondence with half-turns of two helices, forming the double helix (Fig. 5c). On symmetrical level this 3 pairs may correspond to 6 nucleotides structure of telomeres and the characteristics of the structure of the corresponding RNA structure [33]. Upon mapping a polytope onto a polyhedron two points of a polytope are in correspondence with a single point of the polyhedron; hence lifting degeneration in $E^3$ must correspond to the appearance of two more half-turns of two helices. Thus, we get a double helix, where for each turn there is a zigzag union of 6 pairs of elements, not congruent to each other (Fig. 6b). The said 6 pairs form a set of 12 elements with simultaneously six imprimitive 2-element sets (heptagon-heptagon pairs) and 6-element sets (heptagons and pentagons), which is denoted by the symbol 6x2 [13]. The two turns of the double helix are in correspondence with two orbits [6x2, 6x2] where the subgroup $2 \times S_5$ of the group $M_{12}$ is acting.

There is a glide reflection plane $\{m|2a\}_{1/2}$ going through the midpoints of edges of the Petri polygon, making the 6 heptagons contained in them coincide. If one draws a glide reflection plane $\{m|2a\}_{1/4}$ going through quarters of edges of the Petri polygon and parallel to the given plane, it will make the centres of 3 pentagons and 3 heptagons coincide. A union of such polygons, closest to each other, forms a zigzag line out of alternating pentagons and heptagons (grey chain in Fig. 6b) – a pair of a pentagon and a heptagon, multiplied by the doubled translational component $2a$ of the glide plane. The translation by $a$ of the gray chain determines a congruent (up to the rotation of pentagons and heptagons about centres) union of 3 pentagon-heptagon pairs, represented by the white chain in Fig. 6c, and has no "advantage" as compared with the grey chain.

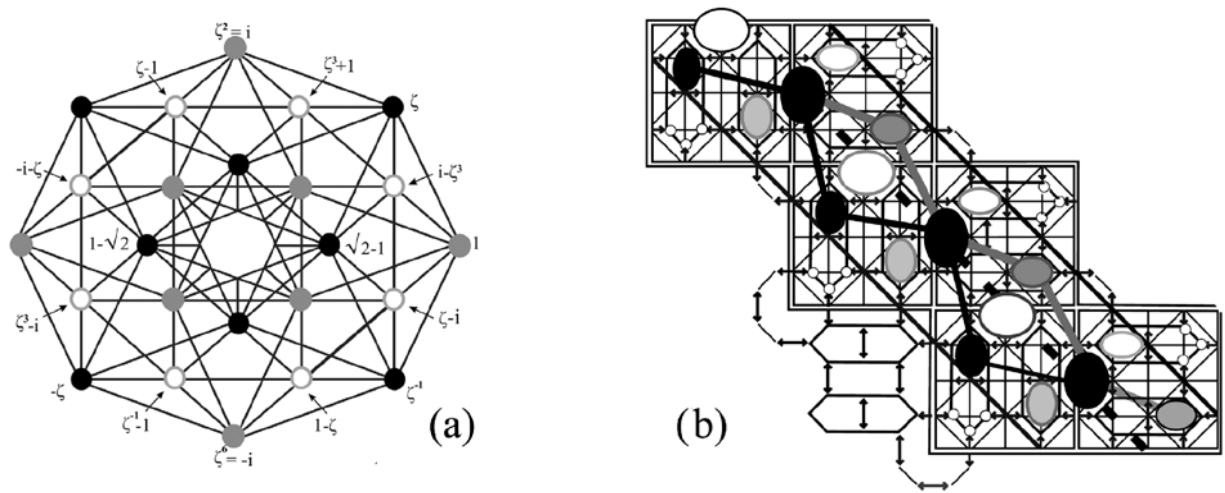

Figure 6
*(a) The mapping of the {3,4,3} polytope onto a plane with the polytope vertices as elements of a nonprincipal lattice (Figure 8.1 in Conway & Sloane, 1988). The colour shows the subdivision of 24 vertices onto three orbits of 8-cyclic group and onto six orbits of 4-cyclic subgroup of the 8-cyclic group.*
*(b) The flat development of the $\{2^2 \cdot 24\}$ polyhedron as 96-vertex subdividing of the flat development of a cube into 5-, 6- and 7-gons. Black (white) ovals single out a zigzag chain composed from three pairs of pentagons and heptagons. The congruent chains are denoted by black - grey and white - (light- gray) ovals. These chains are within the central strip shown by black lines, which corresponds to «equatorial» strip of the polyhedron in Figure 5c. Glide reflection plane is shown by dotted line.*

In a zigzag-like chain of one helix each heptagon is connected to a pentagon and vice versa. An analogous type of connection is preserved also while linking pentagons and heptagons closest to each other from different helices. Summing up the above, it is possible to assert that putting into correspondence with pentagons and heptagons of the flat development of the double helix the syn - and anti- conformations of the bases [15] one obtains a scheme flat development (Fig.7.a) of the Z-form of DNA structure (Fig.7b,c) The transitions between subgroups $M_{12}$: $S_5 \rightarrow M_{10} \cdot 2 \rightarrow 2 \times S_5$ determine the transitions between the corresponding subsets of elements, on which these subgroups act: $[1, 5, 6; 2, 10] \rightarrow [6^2; 2, 10] \rightarrow [6 \times 2, 6 \times 2]$. On symmetrical level these transitions correspond to a junction between B-DNA and Z-DNA [8, 9, 29, 30]. Scheme of this junction is shown on Fig.8.

A union of two half-turns (manifolds) from each helix forms one turn of dabble helix (manifold M) by a standard topological operation of taking a connected sum (#), including such variants of it as gluing on a handle (g) and Mobius film (μ) [14]. In particular, if in the α-helix the assembly is performed by gluing on handles, then in the DNA structures considered below the assembly is impossible to be described without the operation of connected sum proper (Fig. 4b). Application of the said operation plays a special role for the rod structures in question, because it provides an experimentally established possibility to bend them under certain angles [8, 9].

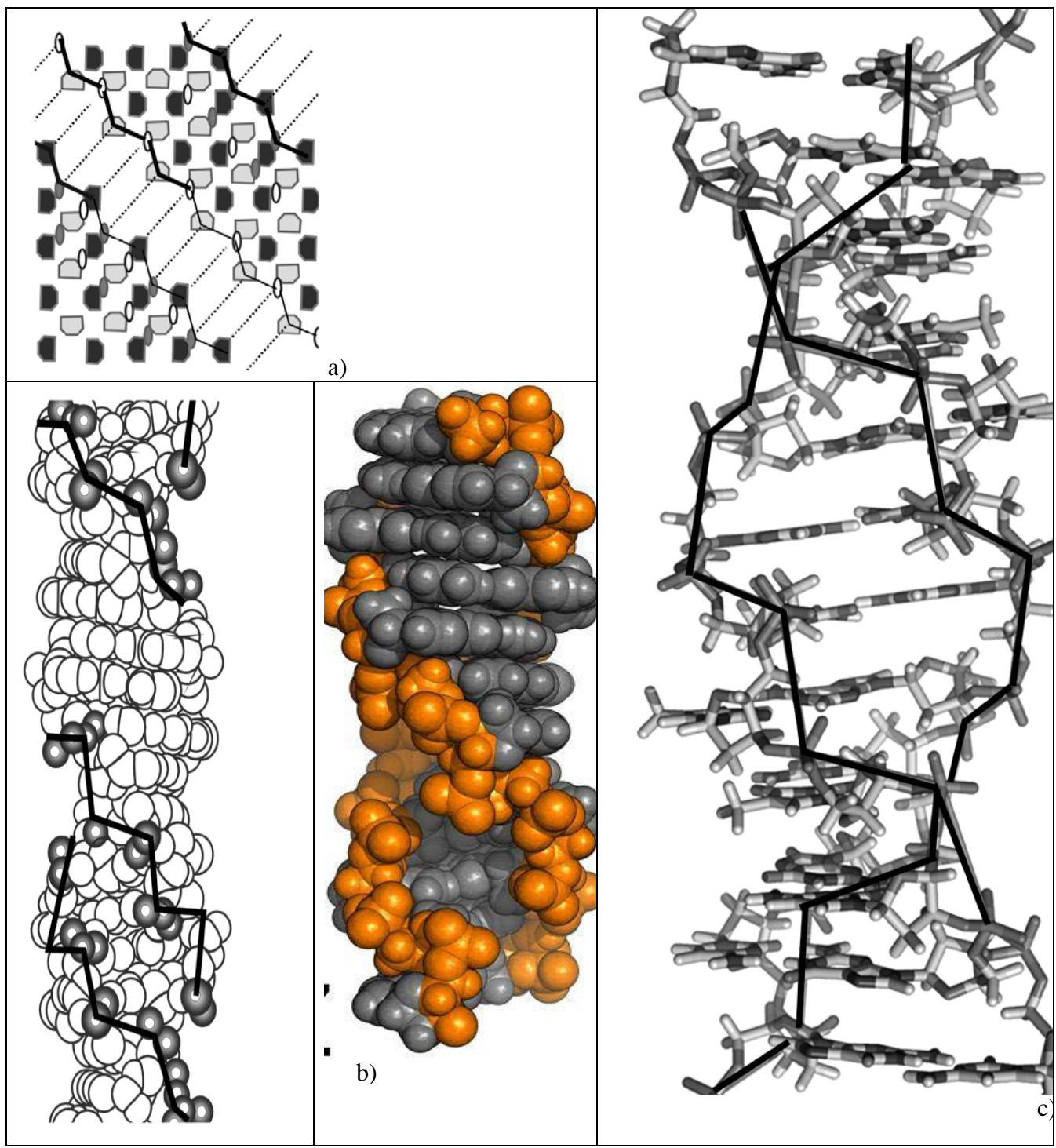

Figure 7
*(a) The cylinder development of double helix as the part of the crystallographic tiling of zigzag chains, which is defined by the central strip of flat development in Figure 6b. Half-turns of helices are shown in thin and solid lines. Zigzag lines determine two left (relative to Figs. 2a, 2b) helices containing six pairs of pentagon and heptagon. In each pair heptagon (pentagon) corresponds nucleotide in a cin-conformation (anti-conformation) in Z-form DNA [15].*
*(b) The repeated unit of the left-spiral Z-form DNA is a two adjacent nucleotide pairs. The spiral rotation angle is equal to $-9°$ or $-51°$ which is dependent on the realization of the contact type (anti-cin-conformations or cin-anti-conformations) in the given point [15].*
*(c) A model of Z- form DNA, zigzagged double spiral is shown by thick black lines (Figure by Richard Wheeler - nickname Zephyris)*

The polytope {3, 4, 3} (Fig. 6a) is a cell of the honeycomb {3, 4, 3, 3}, which may be projected onto a plane in the flat development of the $\{2^2 \cdot 24\}$-vertex polyhedron. According to [16], such a flat development may be obtained from partitioning of the flat development of a cube into 5-, 6- or 7-gons (Fig. 6b). In order to achieve this, the flat development of a cube is embedded into the {4, 4} tiling with an edge 2(a+b) in such a way that the Petri polygon of the cube becomes part of the Petri polygon of the tiling {4, 4}. The edges of the cube may be partitioned into squares belonging to an orbit of the space group {a+b, a-b} 4mm. The normalize of this group is a space group {a, b}4mm, mapping the tiling $\{4,4\}_a$ (with edge a), among whose vertices are the $2^2 \cdot 24 = 96$ vertices of the partitioning of the cube's flat development into 5-, 6- and 7-gons.

A strip of this partition, containing all 6 edges of the Petri polygon of the cube (Fig. 6b), allows one to select a 81-vertex subset, which is in correspondence with 80 vectors out of the 96 mentioned vectors. It can be shown that the given 80 vectors allow one to make transition to a polytope {160} and to obtain the parametric axis 40/11. Tripling the cell $3D_4$ in relation to $D_4$ allows one to finally make a transition to the axis 120/11, tripled with respect to the axis 40/11 in the number of elements (120 is an invariant of the second coordination sphere $E_8$). The polytopes {160}, {480} and {960} [17-20] allows one to use the parametric axis 120/11 in order to characterize some forms of DNA structures.

Simultaneously note that from the viewpoint of algebraic manifolds it is possible to view $S^3$ as $E^3 \cup \{\infty\}$ with subsequent transition to projective constructions, also using subgroups of the Mathieu group $M_{24}$, which is in correspondence with automorphisms of the mentioned Hurwitz group – a starting points in constructing a complex lattice $E_8$. Because the manifold $S^3$ corresponds topologically to SU(2), and the midpoints of geodesics with the intersection of the group SU(2) with the algebra su(2), the role of the 24-element Hurwitz group of unit quaternions, as well as necessity to introduce polytopes become evident.

In the case of homotopic map of $D^1 \times D^2 \to M$, one may use maps of the form $\partial D^3 \to S^2 \to 1$ ($\partial D^2 \to S^1 \to s'_0$), the kernels of the homomorphism in this case are the kernels of root systems, generating reductive groups [22]. Structural features of centralizers of such groups lead to projective groups of the form PSL (2, p). An algorithm of formation for systems in question may be described using various subgroups of Mathieu group $M_{24}$, in particular, the $M_{12}$ group and its subgroups. This group, along with the subgroups $M_{10}$ and $M_{11}$, whose action on the set may be considered using the group $L_2(11)$, contains also subgroups from table 10.3 in [13].

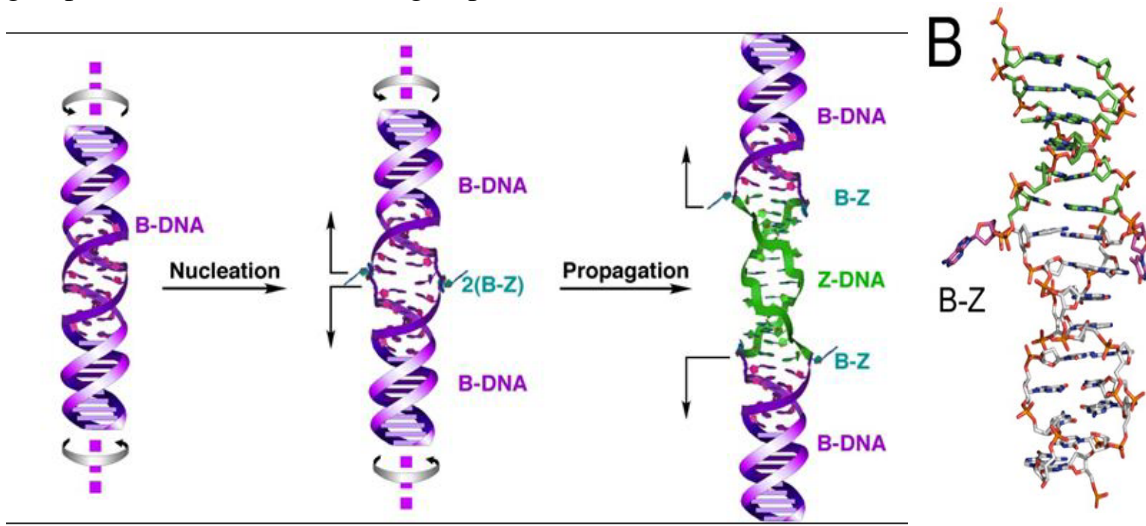

Figure 8. Scheme of a junction between B-DNA and Z-DNA [29, 30]

A connection between subgroups of the group $M_{12}$ (subgroup $M_{24}$, in its turn) such as $M_{11}$, $A_4 \times S_3$ and $4^2 \cdot D_{12}$ describing their action upon appropriate ($L_2(q=11)$) projective manifold ($\Omega = PL(q)$, put into correspondence with a group of linear automorphisms of appropriate vector manifold), establishing an interconnection between them and corresponding sets of orbits of 12 or 11 elements – with selected point ($M_{11}$), a set of orbits of 4 imprimitive subsets of 3 elements plus 3 subsets of 3 elements ([4x3,4x3]), as well as a set represented by 3 imprimitive subsets of 4 elements ([$4^3,4^3$]).

One should also take into account the features of co-representations $M_{12}$, given by relations for elements of transformations that generate $L_2$ (q=11) [13]:

$$\alpha^{11}=\gamma^2=(\alpha\gamma)^3=(\gamma\eta)^2=(\eta\alpha\gamma)^3=\eta^{10} \qquad (5),$$

where $\gamma\delta$ is replaced by $\eta$ and $\delta$ is not considered [13]; presence of characteristic triplets is evident. Then, in the said sense, the $\alpha$-helix, characterized by the parametric axis 40/11, and containing (in cylindrical approximation) 3 or 4 elements per turn, may be viewed as a peculiar matrix for each helix in DNA structures. It is natural that other geometric parameters of molecules that constitute its packing, lead to tripling of the number of elements per turn ($3.6\times3=10.8$) and the pitch. In consideration of two-helix systems such interconnection is not evident.

Upon selecting on a sphere "the north and the south" disks (around the north and the south hexagons) and subsequently gluing the disks, while allowing for a local cylindrical approximation of the minimal surfaces in question, two zigzag-like chains may be put into correspondence with half-turns of two helices, forming the double helix (Fig.7.).

The A-DNA, defined previously, has been obtained by formal tripling of the $\alpha$-helix via transition from the 40/11 to 120/11 axis. In this case, a turn of A - DNA can be present as a result of unwinding of 3 turns of $\alpha$- helix having 11 elements (Fig. 2c). Formally, this is evidenced the value 20° of the deflection angle for conventional coil plane of A-DNA. The presence subsystems of 3 or 4 elements (letters), which correspond to the turns of $\alpha$- helix, is not considered at the turn of A- DNA. Accounting of such subsystems (because of possible disorientations relative to each other), which may consist of three or four letters, are needed for B-DNA Indeed, the presence only three letters is guaranteed that may relate to the triplet genetic code.

Upon mapping a polytope onto a polyhedron two points of a polytope are in correspondence with a single point of the polyhedron; hence lifting degeneration in $E^3$ must correspond to the appearance of two more half-turns of two helices. Thus, we get a double helix, where for each turn there is a zigzag union of 6 pairs of elements, not congruent to each other (Fig. 7a). In the given case there are 21 element per 2 turns, which is 10.5 elements per turn, characteristic for the B form of DNA structure [8].

It is known that insplicing of RNA fragments, corresponding to reading from introns, DNA structures are cut out to be subsequently twisted, as a rule, into ring structures (as a result of the transition from an unstable to a stable configuration), while fragments corresponding to reading from exons are knit together using matrix RNA, corresponding to a certain protein. Correspondingly, it is possible to assume that fragments of exons and introns of DNA structure, as well as DNA structures and RNA parts corresponding to them, are characterized by different parameters, of which the most likely are the laws of one-parameter groups that characterize them, and, consequently, properties of non-integral axes corresponding to such groups. For example, a ring-like structure shown in Fig. 4.d can be characterized by an axis 16/5, when the "joining" of each subsequent 7-vertex complex is realized by reversal of face on $112.5^0$

Constructions containing bifurcation points, whose general features had been considered already by A. Poincare [1], correspond to a general approach known as the catastrophe theory [25]. It may imply the appearance of stable $\alpha$-helix-like systems as well as DNA structures transmitting hereditary features (and, consequently, they ensure both reproduction as well as further evolution of living systems) can be considered in terms of a kind of a catastrophe for the non-living organic world.

An important characteristic of DNA is its virtual double half-turn nature (which defines a "building block" of the double helix as a union of two half-turns of each helix) of the turn of the helix. Worth noting also is a virtual half-turn nature of each turn, as well as a possible role of hidden triple period in relation to using a three-letter basis (for a four-letter alphabet), when it is exactly three nucleotides that determine the type of amino acid in formation of polypeptide chains. The fourth nucleotide cannot be unique due to its structural features, but it is necessary to take into account the role of RNA structures in changes of DNA.

Parameters of the introduced non-integral axes (one-parameter subgroups) may, to certain extent, be viewed as a parameter that characterizes local geometric properties of the matrix "(its lock)", from which the information is being read, and the system of its "key" to which the recording is made. The formation of turns by half-turns from different helices in realization of the double helix structure as a single system, may imply existence of certain cross-exchange (local crossing-over) between the female and male helices (at least, on a considerable part of the helix), whose role is yet to be understood given impossibility of exchanging turns between various helices going in opposite directions and shifted relative to one another.

As for the DNA structure to the configuration types of degeneration than discussed in Section 2 on the example of α-helix, consider the following: the union of surfaces of the helicoid and catenoid, as lifting of degeneracy related to their adjoint nature, double-helix property, as lifting of degeneracy by helical rotation in opposite directions.

## 4. Conclusion

Lattice $E_8$ determines special class of helical axes (30/11, 15/4, 18/5, 40/11 and other [16, 17]) which also contains axes 30/7 = 4.2857; 10/3 = 3.3333; 30/13=2.3077 and 24/11. According to Table 5.1, " the linear groups formed by polypeptide chains " [6] the numbers of residues per turn are +4.3 for π- helix, -3,3 in spiral collagen fibres, -2,3 in twisted β - structure. According to [7], the twist angle of β- strand is 165° = 360°: (24/11). Thus, not only the axis of α- helix, but other axis of spiral substructures of biopolymers (protein secondary structure) determine of *a priori* in the framework of our approach. With the great degree of certainty can be argued that our approach allows to define a symmetry of associations of helical biopolymers forming the tertiary structure of the protein.

Real processes in nature, as a rule, depend upon a large number of parameters, whose number and numeric values are not known precisely. However, while in most cases small variations of parameters transform such systems of specific position into systems of generic position, such considerations are suitable for individual systems, when, for instance, one- or two-parameter systems can, but only for certain parameters values, be transformed by small perturbations into non-degenerative ones. However, if one looks at families of systems, degeneration by small perturbations is impossible to eliminate completely; for example, for one-parameter systems it is impossible to eliminate the simplest degeneracies (given by a single equation in the space of all systems).

The results obtained in the present work (using quite elaborate mathematical constructions) show that an α-helix and various forms of DNA structures are described using methods of algebraic topology as special local lattice packings, limited by surfaces of helicoid-like type. Such surfaces correspond to the bifurcation point for minimal surfaces given by Weierstrass representation and satisfy the condition that the index of an unstable surface equals zero.

Thus, realization of DNA structure (what is said below being completely relevant to the B-form) represents a chain of lifting various forms of configurational degeneracies, namely: a bifurcation point as a non-degenerate point of the Morse function, the union of surfaces of the helicoid and catenoid, as lifting of degeneracy related to their adjoint nature, double-helix property, as lifting of degeneracy by helical rotation in opposite directions. Added also is the use of the real lattice $E_8$, obtained by taking the real part of the complex $E_8$, in order to describe which, in its turn, a 24-element Hurwitz group [13] is used with directional degeneracy lifting (signs of vectors). Furthermore, use of polytopes as homogeneous and isotropic (locally) discrete manifolds with subsequent selection of a bouquet and using the two-valued nature (or one of hemispheres without the pole with subsequent doubling) in constructing stably non-trivial bundles. In fact, as is known [14], the structure of minimal surfaces $M \subset E^3$ is complicated, at the same time, for the case of fixed $S^1 \subset E^3$, there is no theorem for uniqueness of solutions of defining equations. So that the use itself of minimal surfaces, without lifting of degeneracies, related to introduction of locality for a series of geometric properties, reflecting, in particular, homogeneity and isotropy of the physical space (for the scale in question) is of no particular value.

As predicted by a theory of catastrophes [25], formation of such structures corresponds to processes of lifting configurational degeneration, and state stability – with existence of a bifurcation point and other changes associated with the removal of degeneracy of the configuration of various types.

It would be of interest to understand what new perspectives in biophysics of DNA structures can occur (for example, on the part of evolution and functional changes of the genome) when the established topological features of their structure are taken into account, in particular, the stability of such systems and their subsystems, as well as substitution mechanisms for packing elements. Note also such features as an interconnection of local cylindrical nature with frequently used cylindrical approximation; a special role of local lattice property and bifurcations in stability considerations; the necessity to take into account parametric properties of non-integer axes and virtual double-half-turn nature of a turn of the helix. In fact, the two latter characteristics undoubtedly change upon transitions from a state of a unified two-helix system toward separate one-helix fragments, for instance, upon methylation of nucleotide pairs. Finally, it should be established which molecular constructions correspond to special variants of the resulting surface, representing a union of the helicoid and the catenoid in the form of the helicoid wound over the catenoid.

## Appendix 1. Helix and catenoid as minimal surfaces - Weierstrass representations

The groups and algebras used below, as well as the algebras, are related to complete minimal ruled surfaces of revolution like that are possible to describe by one-parameter subgroups. If a catenoid as a closed surface in $E^3$ cannot be extended to a large minimal surface, completeness of the helicoid as a minimal surface without an edge, ensures, in fact, that it is maximal in a sense that any extension of its fragments to a greater minimal surface remains a domain on a helicoid (preserving in combination with an exterior metric the property of the entire embedding space being metric). Using bandings that preserve a class of minimal surfaces, it is possible to wind a helicoid over a catenoid, so that there is a one-parameter family (associated surfaces) of isometric minimal surfaces (with smooth dependence upon a parameter) from helicoid to an infinite-valued winding of the catenoid (adjoint surfaces) with a metric-preserving diffeomorphism existing between them.

The associated surfaces thus obtained for various angles $\alpha$ in the form of deformed helicoid are shown in fig. 26 in [3]; at the same time, with increasing $\alpha$ the curve describing a tentative sum of vector radii forming the helicoid and the catenoid will increasingly veer off from the curve corresponding to a cylindrical surface. Changing (decreasing) the helical pitch [3], it is possible to obtain that a film stretched over a contour consisting of two helices and two closing segments, will cease to be a helicoid and turn into a surface shown in fig. 27 in [3]. Appearance of the bifurcation point for a catenoid for certain ratios of radius and pitch allow one to preserve local cylindrical properties for the surfaces in question. The constructions mentioned above allow one to begin using global Weierstrass' representations.

It is shown [3] that Weierstrass representations allow one to define catenoid as well as complete helicoid, and, in general case, an associated family for some minimal surface M (for instance, of helicoid or catenoid) consists of locally isometric minimal surfaces (incongruent pairwise, as a rule). The said properties allow one to find an adjoint family of surfaces unifying the helicoid and the catenoid, given by a radius-vector

$$r(u, \varphi, \alpha) = r_1(u, \varphi) \cos\alpha + r_2(u, \varphi) \sin\alpha \qquad (A.1),$$

where $r_1$, $r_2$ are radius-vectors, describing the catenoid and the helicoid, $(u, \varphi)$ are coordinates on the surface, and $\varphi$ corresponds to the angle in cylindrical coordinates. The parameter $\alpha \in [0, \pi/2]$ is such that for $\alpha=0$ we have a winding up of the catenoid, and for $\alpha=\pi/2$ – a helicoid, and the angle of inclination of the turn is in correspondence with a conditional one, with respect to plane perpendicular to the Z axis (Figs. 24, 25 in [3]). For some fixed value of the angle $\alpha$ it is possible to build a construction that is a "sum" of a catenoid with radius $\cos\alpha$ (the parameters above are normed for a=1 and h=2π) and a helicoid with distance between turns equal to h=2πsinα. Note immediately that it is possible to give an estimation for the said turn in the point of bifurcation, if one uses the relation between these parameters, given in (A1), which leads to the value tgα=0,382 and the angle α≅20,8$^0$.

A catenoid determined by one such surface is locally isomorphic to a helicoid belonging to a class of minimal surfaces allowing for conformal coordinates. Helicoid may be represented as an infinite – valued surface over a catenoid (folds remain within the class of minimal surfaces) or onto a sphere without poles [3]. Correspondingly, if we have an axis formed by two circumferences sharing an axis (they are of the same radius and are situated in parallel planes), then in order to describe all minimal surfaces closing in on the contour it suffices to describe all catenoids closing in on it (Fig 1b,c).

Making a significant simplification, we shall assume that the relation (A1) determines a condition, rather important for applications, for the transition from a locally minimal to a locally cylindrical surface, namely, the surface, for which the neighbourhood of every point is approximated by a cylindrical surface. In a bifurcation point (non-degenerate for a Morse function) a topological regularity is broken [14]; at the same time a cell structure considered in the preceding section must appear on the manifold. For some helicoidally similar structures the parameters of a bifurcation point may be determined in an approximation of their cylindrically similar surface.

In diamond-like polytope $\{q(2^\gamma \cdot 24)\}$, q is an integer, $\gamma = 0,1,2$, $\gamma=0,1,2$ the union of $2\{q_1\}$ vertices on 2 neighbouring circles $S^1$ forms a Q-chain, similar to a <110> – chain in a diamond structure. The maps of a Q-chain into a Q – edge we shall denote by two-headed arrow. A 3D rod substructure of a diamond-like algebraic polytope – the rod K corresponds to a face of the loaded polyhedron $\{2^\gamma \cdot 24\}_q$ (Fig. 5). Upon mapping it onto a sphere it corresponds to a union of 4 channels 30/11; and in the interstices between them 3 channels 40/9 appear (Fig 5a.). For a surface M, given by a Weierstrass representation (C, f dw, g), the index Ind M is finite if and only if the function g is fractional - rational [3]. The symmetries 30/11 and 40/9 are in correspondence with such functions ([20]; hence, the minimal surfaces corresponding to such non-integer axes will possess finite Ind M, and some degree of the topological stability.

Let M be some surface given by Weierstrass representation (U,w, (aw+d)$^m$)), a,b∈C, a≠0, where m is a non-zero integer, U⊂C is some subdomain of the complex plane (as a result of complexification for discrete constructions). The surface M is characterized by zero index if the image of the domain U under a Gaussian map is contained in some open subset W of the sphere $S^2$ [3]. The subset W can be defined either as an open hemisphere of the sphere $S^2$ without the pole, or as the part of the sphere $S^2$ (about 5/6 of the total area), contained between two parallel planes which are separated from the center (zero) of the sphere by the distance th $t_0$, where $t_0$ – is the only root of the equation cth $t_0=t_0$.

Polytopes are extreme of the volume functional, including also locally minimal manifolds. For a surface M, given by a Weierstrass representation (**C**, f dw, g), the index Ind M is finite if and only if the function g is fractional - rational [3]. The symmetries 30/11 and 40/9 are in correspondence with such functions [3]; hence, the minimal surfaces corresponding to such non-integer axes will possess finite Ind M, and some degree of the topological stability. For the polytope $\{12(2^\gamma \cdot 24)\}$ given by the vectors of the 2$^{nd}$ coordination sphere of the lattice $E_8$, the subset of 5/6 of its vertices determines the polytope $\{10(2^\gamma \cdot 24)\} = \{12(2^\gamma \cdot 20)\}$, which can be mapped into polyhedra $\{2^\gamma \cdot 24\}_q$ and $\{2^\gamma \cdot 20\}_q$, determining the possibility of existence of a surface with a finite (and possibly zero) value of Ind M.

These representations, but using a pair of a holomorphic 1-form (ω) and a meromorphic function (g), belong to global ones; then minimal surfaces containing irregular points, are called generalized minimal surfaces. Such definition of a surface implies that local vanishing of Gaussian curvature, for example, for discrete umbilical points that also are zeros of the function g in Weierstrass' representation on a regular surface, does not in any way imply that the surface's mean curvature equals zero. It is appropriate to recall here that [3], if two minimal surfaces (embedded in $E^3$) touch in some interior point P, and, at the same time one of them is locally (in a neighborhood of P) is to one side of another, the neighborhoods of the point P (which may include vertices of polyhedra, lying on a surface, and it is possible to put them in correspondence with midpoints of geodesic lines лежащие на поверхности, которые, в свою очередь, можно соотнести с серединами геодезических, situated, in contrast to polyhedral edges, on a surface and joining these vertices) are locally planar. In the transition to the vector representations of the plane, in fact, is used, the field of complex numbers is a two-dimensional vector field over the field of real numbers, containing, as it is known [14], the field of rational numbers (which, unlike the algebraic numbers are not conjugated, non-themselves).

In what follows, while considering surfaces with finite (in particular, with zero instability index under additional conditions) index, when the function g belongs to the class of "good" [3], in particular with finite number of zeros, is defined using polynomials (and an arbitrary holomorphic function) as well as the requirement of its fractional rationality, appears a possibility to put into correspondence zeros of the so-called p-polynomials, for which, in what follows, we use the basis of polynomials (not containing squares) of Weyl group of the root lattice $E_8$, which, in facts, represents a realization in the form of the lattice of zeros of such polynomials. In fact, we are using the theorem that for a vector group and a G-closed subgroup of co-dimension 1, there always is a p-polynomial, for which G is the set of its zeros [22].

Absence of common roots, realized due to polynomial degrees being relatively prime or via their expansion into primes, ensures that the requirement above is satisfied. Thus, one ensures consideration of surfaces with finite number of singular points, defined using local characteristics (invariants as well as closed sets of vectors, whose isomorphism's with e-groups (exponentials of p), in particular, with unipotent or one-parameter groups (in the form of using exponential elements) allows one to consider isomorphism's of the form $G \to G_\alpha$, or $T \to G_m$ for a torus as characters of representations, because additive and multiplicative groups $G_\alpha$ and $G_m$ are unique (up to an automorphism) one-dimensional algebraic groups of the root lattice $E_8$, as a kind of pre-phase.

When using global isothermal coordinates for helicoid systems (u,v), the Cartesian coordinates of points can be expressed via hyperbolic functions, namely x=shu·chv, y=shu·shv, z=v. For this consideration it is essential that in the case of such transformation the coordinates are related to the conformal ones, and any conformal transformation on a sphere that is close to identity both in $E^3$ as well as in $S^3$ ($S^2$) may be represented as a one-parameter of the form exptA. If the tensor of deformations (measuring distance distortions) equals zero, then from vector fields giving conformal transformations it is possible to turn to using corresponding algebras (subalgebras), giving motions. The point is that for oriented surfaces, transition functions (coordinate transformations) give a complex structure, so that the pair u,v is replaced by the complex coordinate z=u+iv.

If a minimal surface contains singular points, it belongs to generalized minimal surfaces. Weierstrass' representation used below to define minimal surfaces, describes local structure of minimal surfaces. Such representations, corresponding to various domains on the surface, can be glued into a global Weierstrass representation, which should be viewed as a set of topological spaces, glued from coordinate domains (this is realized essentially for cell complexes).

## Appendix 2. Lattice $E_8$, polytopes, one parameter subgroup of symmetry group of polytope and non-integer axes

The interest to semisimple groups and, in particular, their algebras is related to the following important circumstances, namely, as shown in [1, 23], when considering manifolds with or without an edge, wave fronts and caustics (the latter defined as an envelope of a family of geodesics (for a metric of general position), and, consequently, features of reflection groups, functions on them, their representations, as well as geodesics and projections onto a plane), typical are the features of Weyl groups of simple groups of types $A_n$, $B_n$, $D_n$, $F_4$, $G_2$, $E_6$, $E_7$, $E_8$ and their algebras. Introduction of local lattice properties allows one not only use the fiber bundle formalism, and, in particular, covers, but also preserve definition of coordination polyhedra. Simultaneously, a possibility appears to use not just invariants (indices) of the root lattice of the special algebra $e_8$, and consider manifolds of non-regular orbits (MNO) for non-crystallographic groups, generated, for instance, by $E_8$ and viewed as subgroups of transformation groups of Euclidean vector spaces.

In fact, consideration of the root lattice $E_8$ of special algebra $e_8$ allows one to solve several problems. First, for a unit sphere one may take the first and next coordination spheres of the $E_8$ lattice using both sets of root vectors to define vector manifolds, as well as invariants of such lattice, along with relative invariants characteristic of the mentioned spheres. Such discretization, as well as using finite groups and Weyl groups simplifies the use of an algebraic approach. Furthermore, it is possible to use the fact that topologically $S^3 \cong SU(2)$, so that application of polytopes defined on $S^3$ and generated $E_8$ as homogeneous spaces, then using local homogeneous manifolds, allows one to consistently remain in the realm of local algebraic approach. Note that consideration of one-parameter subgroups $SU(2)$ is reducible to studying a Cartan (maximum commutative) subalgebra, and for geodesic manifolds (on a disk, for example) – to Grassmanian manifold $G^C_{2,1}$.

In the case of finite reflection groups [23] the characteristic equation (where the product is taken for $\nu=1\ldots n$) has (for odd n) the form $\prod_\nu (\lambda - \exp 2\pi m_\nu/h)=0$, where the characteristic root has the following form $(\exp 2\pi/h)^{m_\nu}$. Then the complete number of reflections and the order of the group ($\Gamma$) will be determined by invariants, namely, $\Gamma = (m_1+1)\cdots(m_n+1)$. Thus, the character (representation) is defined as a finite Fourier series with non-negative integral coefficients, which for finite reflection groups are reduced to exponentials or invariants (see table in [4]). Because an expansion (into exponentials or invariants, if one add up by 1 for identity transformations) into similar coefficients for a root lattice, for example, $E_8$, are known, it is not hard to pay attention to the parametric meaning of the non-integer axes (m/p) being used, allowing one to consider discrete systems, given as homogeneous spaces in the form of polytopes. Polytopes are the extremes of the volume functional, which include locally - minimal varieties. Properly, the transition from consideration of the polytope on $S^3$ to using constructions of the form $S^1 \cup S^2$, leads to a necessity to define on $S^1$ certain algebraic constructions with p elements. In particular, these constructions must be in correspondence with invariants for 4D root lattices (the greatest exponential being equal to 11 for the system $F_4$ whose roots correspond to the roots of the first and the second coordination spheres of the lattice $D_4$), that is lead to introduction of additional intermediate conditions of the form $(\exp 2\pi/p)^p=1$ which allow one to introduce periodicity.

In this case the centers of cell complexes (clusters) form a system of points on a helix, which may correspond to a 2D lattice on the flat development of a cylinder, determined by one of the axes of Gosset's helicoid L/p:

$$L/p = 2^\gamma \cdot 8I_n / 4k_{js}m_{js} \quad (A.2)$$

where $2^\gamma \cdot 8I_n$ and $8I_n$ – are the number of vertices from the 2nd and the 1st coordination spheres of the lattice $E_8$; $\gamma=0,1,2$; $I_n$, $I_s=k_{js}(m_{js}+1)$ are invariants of $E_8$, $k_{js}$ are integers, $m_{js}$ is the index of a lattice embedded in $E_8$ [16-21]. The value L/p, where p is an integer or some power of it, corresponds to an exponential representation $\exp 2\pi ip/L$ or one parameter subgroup of symmetry group of polytope. If one uses exponential representations which will be in correspondence with the rotational part of elements, describing the motion of the point along the curve while using the vectors mentioned above, then the original invariants will be taken as well as their expansions into primaries. In fact, the non-integer reference axis is defined by the elements of the reflections of Coxeter.

Each finite reflection group possesses a finite basis of invariants [23] that can be identified both with a set of coordinate functions as well as with degrees of their invariants. For illustration it is convenient to list exponentials p (invariants are given as p+1) for the systems in question: $E_8$ (1,7,11,13,17,19,23,29), $H_4$ (1,11,19,29) and F4 (1,5,7,11), in order to select one common to all of them – 11, and also for the group группы $H_3$ used to introduce a parameter into $H_4$ (1,5,9) and $D_4$ (1,3,3,5) – in order to select common exponential (6 being the appropriate invariant). The selection itself of one or the other invariant does not by itself define necessary constructions, as it is done below, but serves as a "control" indication. The point, in particular, is that a transition to finite discrete manifolds is not possible without introduction of finite groups and invariants that characterize them. For example, in derivation of crystallographic groups one uses the fact that all discrete subgroups of the orthogonal group are finite point groups (crystalline classes).

Note that in the case of symplectic groups any invariants ($\Omega$) are expressed via combinations dependent upon appropriate p – pairs of vectors (covariant and contravariant), at the same time, in a n-dimensional vector space there are invariants ($\Omega_\alpha$) corresponding to partitioning the number p into a sum of odd distinct summands $p=h_1+h_2+...+h_\alpha$ for $h_1<h_2...<h_\alpha=1$ (mod2) ($0<h_\alpha<2n$), which form a basis for linear invariants of degree p of the adjoint group.

Thus, along with expansions of the form like 40=23+17=29+11, it is possible to realize less stable or unstable systems corresponding to expansions containing invariants as well as just odd numbers (40=19+21) or consisting only of odd numbers (given that the condition above holds true) that are not simple (15+21+39+45=120 when 120 is viewed as a relative invariant of the 2 – coordination sphere $E_8$).

A tetrahedron approximating the packing of 4 amino acids is a simplex in $E^3$, the union of tetrahedra determines a simplicial complex [14]. According to [24, 26-28], if the chains of one type form the rod K, it is an orbit of a screw axis, combining a rotation by the angle L/p and a shift by the vector h along the axis. Such a channel may determine a topologically stable (with come finite Ind M) helicoid-like structure $\Omega$ which satisfies the relations:

$$\Omega \leftarrow <L/p \mid \lambda\, h(\mathbf{r}(u,\varphi,\alpha))> 2\{q\} \leftarrow \{q\,(2^\gamma\,24)\} \leftarrow E_8 \qquad (A.3),$$

where $\lambda$ is an integer, $\gamma = 0, 1, 2$. The radius determined by $r\,(u,\varphi,\alpha) \approx h/(2.4)$ (3) represents the radius of a cylinder-like surface, into which $\Omega$ can be mapped. The angle of rotation of the axis L/p depends only on invariants of $E_8$; for instance, substituting in (A2) $I_n = 20$, $k_{js}=2$, $m_{js}=11$, $\gamma=1$ determines the axis 40/11. At the same time the local lattice property is mapped depending on the direction of the rotation axis, as well as the magnitude h from the value $\mathbf{r}(u,\varphi,\alpha)$.

## Appendix 3. Singular points on minimal surface, simplicial complexes, features of space curves

If a minimal surface contains singular points, it belongs to generalized minimal surfaces. If its Gaussian curvature equals zero, then these points are exactly the zeros of Weierstrass' representation. The second fundamental form of the surface itself may be viewed as a normal component of acceleration vector for the curve $\gamma(t)$ on such surface. Weierstrass representation describes local structure of minimal surfaces. Such representations corresponding to various domains of the surface, can be glued together into a unified global Weierstrass representation, which should be viewed as a set of topological spaces, glued from coordinate domains (in a way this is realized for cell complexes).

All generalized orientable minimal surfaces in $E^3$ (embedding and submersion in $E^3$ are locally equal) are described by global Weierstrass representation, in order to define which it is necessary to define and integral of 1-form on the surface M along a piecewise-smooth (smooth) curve $L \in M$. Then the zero-index of such a surface (as well as the coefficient Ind) correspond to zero dimension of the kernel of bilinear fundamental form (which also holds when defining on a torus).

The second (as well as the first) fundamental (differential) form may be expressed via elements of appropriate algebra and the mentioned forms. In fact, every p-form may be viewed as linear and defined on some vector space, and each element of such space as a linear function on a space of p-forms. Correspondingly, 1- and 2 – forms may be viewed as vector-dependent. Considering singular points on a minimal surface with zero curvature, for integral values of torsion to be not identically equal to zero, it is necessary that the curve (an edge is going to be minimal in $E^3$ when considering complexes, and not minimal – within the surface metric – geodesic line may lie on a surface) joining the mentioned points does not lie on the given surface.

Therefore, applicability of standard definitions of knots for curves lying on conditional strips raises doubts in some cases, if a selected strip is not described by a single curve related to a screw, and the strip itself does not correspond to its dense winding over a cylinder. Moreover, selection of the said strip for packings is hardly of high value.

As is known [3], cyclic Steiner nets are not realized on a plane surface, and, therefore, the sum of edges, joining selected points on a surface under study, does not form a minimal net. However, the situation is simplified within a local approach, when from consideration of a disk $D^2(\partial S^3 \cong D^2)$ it is possible to move to using a torus corresponding to it, as well as plane torus $T^2$, as a local analogue of transition from consideration of a manifold given on $S^2$ to homotopic manifolds on a plane. In such case, taking into account local lattice properties, it is possible to use a standard classification of closed minimal nets on a plane torus [3], because plane tori are isometric (similar), if the lattices generated by them are also isometric

The classes themselves of such tori are in one-to-one correspondence with a number of such lattices. While there are only 10 non-isometric closed minimal nets that can be defined on $S^2$, minimal nets on a plane torus $T^2$ may be deformed without change in their topologies (of which there can be infinitely many). In fact, isometry of such tori is ensured by the condition of equality of lattices generated by them (using the converse condition in order to preserve isometry within local lattice structure being established). Because disks are being used in order to define appropriate tori (to be more precise, manifolds on them) as cuts of the root lattice $E_8$, put into correspondence with automorphisms of the said lattice, the logic of constructions being used leads to the group $G_2$, its root lattice $A_2$ (the first and the second coordination spheres) as well as to Chevalley groups of type $G_2$ to consider appropriate exponential representations.

As is known [1], geodesics on a Lie group, in particular, for the group $SU(2) \cong S^3$ being used, include all one-parameter subgroups and their shifts by arbitrary elements of the group. Under the condition of introducing an exterior metric (when for real manifolds one fixes types, and, consequently, bond lengths), critical points of the area functional realize analogous points for a volume enclosed by such a surface. In building the algebraic constructions being described, one uses disks $D_0^2$ (as plane sections $S^7$ by a three-dimensional plane going through the origin of the principal bundle for $SU(2)$, namely, the lattice $E_8$ for discrete constructions being used); at the same time, the requirement that local minimality is not broken (as well as minimal geodesic property of the disk) is reduced to that all rotations of the disk around its boundary $S_0^1$ must be realized by interior automorphisms of the group, which for one-parameter subgroups leads to Chevalley groups of type $G_2$. Upon introducing $S_0^1 \subset T^1$ as a part of maximum torus in the group $SU(2)$, invariants of Weyl group of type $E_8$ (whose root lattice is considered when restricting to the sphere $S^7$) may be put into correspondence with parameters $D_0^2$.

Let us consider constructions providing compatibility between a minimal surface with discrete structure (which it bounds) and spatial curves connecting singular points of the surface – the vertices of the given structure. Transformation groups for surfaces or bodies in Euclidean space can be considered as manifolds; then, a transition from using the mentioned groups to using ordering automorphisms of systems does not change the situation. The space (manifold) $CP^1$ may be obtained from the sphere $S^3$ by identification of each circle $S^1 = \{\exp i\varphi, 0 \leq \varphi \leq 2\pi\}$ with point of $CP^1$.

In relation to the fact that provisional surfaces and axes to them are considered, introduces in order to describe a turn, generally not lying on a plane, let us introduce a Darboux vector, corresponding to the local approach being used. Darboux vector is the vector showing direction ($\delta$) of the immediate (local) axis of rotation, around which the trihedron (unit vectors of the main normal, tangent vector, and the binormal b = [v, n] to the curve L) for the curve L rotates under uniform motion of a point by L. Darboux vector lies in the rectified plane of the curve L and is expressed via unit vector of main curvature (n) and tangent **t** to the curve L:

$$\delta = (\kappa^2+\chi^2)^{1/2}(\mathbf{t}\cos\theta+\mathbf{n}\sin\theta) \qquad (A.4)$$

where κ и χ are curvature and torsion for the curve L (for example, corresponding to the edge or middle of the strip) and θ the angle between δ and L [14].

As is known, such parameters of a spatial curve in Euclidean space as curvature ($\kappa$) and torsion ($\chi$) comprise a complete set of its geometric invariants, and if a curve is given by vectors, for example, by an orthonormal basis for speed of motion along a curve and normal and binormal vectors, under certain conditions, the latter two may be defined with values for a basis in an appropriate subalgebra. Namely, in order for relation of the form $\kappa = c\chi$ to exist between mentioned parameters, where $c$ is a constant, it is necessary that there is such a vector u that <uv>=const (here v=dr/dl=$k$n, where n=$d^2r/dl^2/|d^2r/dl^2|$). The equations $k=k(l)$ and $\chi=\chi(l)$, where l is a natural parameter [14], are called natural equations of the curve.

*S*ingular points of this surface are related by transformations given by a special homogeneous manifold - an "algebraic" polytope generated by a subsystem of the root lattice $E_8$ of maximal exceptional Lie algebra $e_8$, which allows one to define a homogeneous manifold, bounded in space as well as in number of elements [16, 17, 18]. Using the $E_8$ lattice is explained not only by its being an octonion lattice which is at the top of the row of possible "numbers": real – complex – quaternions – octonions. [13]

The lattice $E_8$ determines both the polytope {3,3,5}, as well as the polytope {240} discovered by Coxeter – a diamond-like union of the two polytopes {3,3,5} [4,10] on the 3D sphere $S^3$ (Fig. 1a, b). The polytope {240} starts the sequence {q($2^\gamma \cdot 24$)}, q is an integer, $\gamma = 0,1,2$, of algebraic polytopes, determined by the 2$^{nd}$ coordination sphere of $E_8$ .[19,20]. The choice of the origin in a deep hole of the $E_8$ lattice determines the sequence of coordination spheres of 16, 128, 448, and 1024 vectors [13]. This allows one to select a subset of 1152=128+1024 vectors of the 2$^{nd}$ coordination sphere $E_8$, corresponding to the polytope {1152} = {12($2^2 \cdot 24$)}, whose substructures are, in fact, used in subsequent constructions.

This subset of vectors corresponds to the polytope {1152} ={12($2^2 \cdot 24$)}, whose substructures, inparticular3Dcuts,goingthroughthecentersofthesphere$S^7$ (principal bundle for the group Su(2)) of the lattice $E_8$in the form of manifolds on disks $D^2$, are, in fact, used in further constructions. Introduction of local lattice structures allows one to not just use the formalism of fiber bundles and covers, in particular, but also to preserve the definition of coordination polyhedra. In fact, considering the root lattice $E_8$ of the special algebra $e_8$ allows one to solve several problems. Firstforaunitsphereonemayusethefirstandsubsequentcoordinationspheresinthelattice$E_8$, using sets of root vectors in order to define vectors manifolds as well as invariants of such lattice, along with relative invariants characteristic of the said spheres. Such discretization and application of finite groups and Weyl groups simplifies the use of algebraic approach. Further more it is possible to use that topologically $S^3 \cong SU(2)$, so that the use of polytopes defined on $S^3$and generated by $E_8$as homogeneous spaces with subsequent use of locally homogeneous manifolds, allows one to remain consistently within a local algebraic framework.

If a minimal surface contains singular points, it belongs to generalized minimal surfaces. Weierstrass' representation used below to define minimal surfaces, describes local structure of minimal surfaces. Such representations, corresponding to various domains on the surface, can be glued into a global Weierstrass representation, which should be viewed as a set of topological spaces, glued from coordinate domains (this is realized essentially for cell complexes).

The latter allows one not only to select toroid subsystems but also one-parameter subgroups characteristic of them, which can also be considered using parameters of the non-crystallographic axes defined above, as well as exponential representations giving the angle of helical rotation (by analogy to how this is done for screw crystallographic axes.

As is known [14], geodesics on a Lie group, in particular, for the group SU(2)$\cong S^3$ being used, include all one-parameter subgroups and their shifts by arbitrary elements of the group. Under the condition of introducing an exterior metric (when for real manifolds one fixes types, and, consequently, bond lengths), critical points of the area functional realize analogous points for a volume enclosed by such a surface. In building the algebraic constructions being described, one uses disks $D_0^2$ (as plane sections $S^7$ by a three-dimensional plane going through the origin of the principal bundle for SU(2), namely, the lattice $E_8$ for discrete constructions being used); at the same time, the requirement that local minimality is not broken (as well as minimal geodesic property of the disk) is reduced to that all rotations of the disk around its boundary $S_0^1$ must be realized by interior automorphisms of the group, which for one-parameter subgroups leads to Chevalley groups of type $G_2$. Upon introducing $S_0^1 \subset T^1$ as a part of maximum torus in the group SU(2), invariants of Weyl group of type $E_8$ (whose root lattice is considered when restricting to the sphere $S^7$) may be put into correspondence with parameters $D_0^2$. Actually used, that the field of complex numbers is a two-dimensional vector field over the real numbers.

A transition from $S^3$ to a bouquet $S^1 \cup S^2$ with subsequent definition on them of vector (projective) finite manifolds (tori) has certain topological features related to existence of two diverse homotopic classes for the map $T^2 \to RP^2$ ($RP^2$ being a non-orientable surface), which not only leads to doubling (two-valued character), but also allows one to consider, when selecting from $S^3$, toroid-like constructions in the form of a catenoid with two circles.

Because Weierstrass representations are used when surfaces are defined (where meromorphic functions are used), an opportunity appears to consider special classes of bundles as level functions of such a function f: $M \to CP^1$ (M- being a 2D compact complex manifold), where the set of maps of singular (critical) points (outside a vanishing cycle [14]) consists of locally identical elements. It is in such a case, that bundles are realized from pairs of domains, corresponding to clockwise and counter clockwise helical rotations.

## Appendix 4. Features and relationship of functional of area and volume, morsovisation of problem of finding their extreme.

Various isomorphism's of homotopy groups allow one to go from consideration of the length functional to multidimensional problems, in particular, to consideration of the area functional ($A_f$) as well as Dirichlet functional $D_f$ (an analogue of one-dimensional action functional of the two-dimensional case). A delicate point is that in a one-dimensional case minimality of the trajectory (geodesic line) automatically implies its smallness, and vice versa (for Morse functions we have a requirement of piecewise smoothness), while in the two-dimensional case smallness of $D^2$ does not imply its geodesic property. In the case of maps in generalized conformal coordinates of a two-dimensional minimal surface in $E^3$ we have, in particular, $A_f = D_f$, so that, given the conditions given in 1 for discrete variations of the disk we have common extreme for both functionals.

Moreover, it is possible to consider maps, distinct from harmonic ones, and at the same time the area functional may remain unchanged, while Dirichlet functional is not necessarily conserved (in general, $A_f \leq D_f$). If the functional is considered in the space of 2D disks, then a unitary periodicity appears [14], which is reduced to the theorem that $\pi_{i-1}(SU(2m)) \cong \pi_{i+1}(SU(2m))$ for $1 \leq i \leq 2m$, which allows one to transition from consideration of functionals (of length or action), defined on a trajectory in a unitary group (SU(2) is used below), namely, the data concerning classes of map of the type $S^i \to M(x_0)$ ($x_0$ being a fixed point from M), given by elements of reduced homotopic groups, to construction of maps considered below for spheres of various dimensions as well as the bouquet $S^1 \cup S^2$. Note that local minimality, used in what follows, implies that volume of the subset does not change in first approximation (first derivative equals zero) under infinitesimal perturbations.

More complicated is the situation with volume functional and with transition from elements of one-dimensional homotopic groups to three-dimensional ones. Submanifolds of zero (identically) mean curvature in each point (locally minimal) correspond to volume extreme, in the sense that it is not changed by infinitesimal variations, while for finite variations volume decreases (introducing exterior metric, this is not directly relevant to our consideration). Note that any completely geodesic manifolds belong to locally minimal ones (with second fundamental form equal to zero). The problem is to find minima (relative ones) of the volume functional in each homotopic class with regard to possibility of appearance of non-removable strata with change in dimension. However, it has been proven [1] that there are globally minimal surfaces (for which there is no problem of non-removable strata), that realize absolute minima of the volume functional, present in each homotopic class. A general approach (moreover, there is a possibility to realize such surfaces when one finds minima that allow for introduction of a continuous parameter independent of change in construction) is to define on the manifold M a stable non-trivial vector bundle (realized in the present approach when non-zero vector fields are given on a torus) then within a restriction of treatment to such a bundle, it is possible to find a globally minimal surface $X \in M$.

A general approach is based on existence of a connection between the number of stationary (critical) points of Morse function of compact manifolds and their diverse invariants, such as characteristics of homology (cohomology), Eulerian and others. For real physical systems it is essential that local domains containing a critical point are associated with special properties of fluctuations, for example, unusual (as compared with regular points) dependence of the average square of density fluctuation vs. domain volume) [1]. Furthermore, in periodically ordered systems, highly symmetric (of high index) points are associated with domains of energy concentration (accumulation), for example, of electromagnetic vibrations.

A restriction to semisimple algebras allows one to consider various functionals (F) as integrals of the system (for example, energy or Dirichlet functional, determined by a quadratic form) as a Morse function. Given a cover on $S^2$, as a family of domains $\{U_\alpha\}$, so that the form $\Omega$ is exact on each $U_\alpha$ (as closed forms on $S^2$) and one may define a map $M^n \to S^1$, then integer cogomologies of the functional F give rise to cell complexes. Given a cover on $S^2$ defined as a family of domains $\{U_\alpha\}$, so that the form $\Omega$ is exact on each $U_\alpha$ (as closed forms on $S^2$) and one may define a map $M^n \to S^1$, then, as it has been pointed out, integer cogomologies of the functional F give rise to cell complexes. A special case is represented by so called dynamical systems – smooth vector field tangent to the manifold M, in particular, dynamic systems on a torus. For such systems, each rotation or map $\phi$: $S^1 \to S^1$ is in correspondence with diffeomorphisms of the form: $\phi(x+1) = \phi(x) + 1$, so that $\phi^n(x_0) = x_0$ for some integer n with the rotation angle $\alpha = 2\pi p/n$ and $\phi^n(x_0) = x_0 + p$ for some integer p. Rationality of the number of rotations in ensured by $\phi$ possessing a periodic point, and if the number of rotations is described by an irrational number, then the number and position of points on the circle $S^1$ corresponds to the angle of rotation $\alpha$ for algebraic constructions $\alpha = 2\pi p/h$.

In particular, in the domain of phase transitions (which most often accompany evolutionary processes), as is known [1], there are critical points; at the same time, mathematical definitions of those points have certain and adequate physical content. Indeed, for smooth (continuous) functions (f), the number of degenerate singularities is not large, while non-degenerate singularities, under certain conditions [1], may merge with each other, giving degenerate ones. The critical points themselves are subdivided into degenerate and degenerate ones depending on the absence or presence of non-degenerate metric of second derivatives which characterize the given point. At the same time, $d^2 f$ is a scalar bilinear form on the tangent space and does not depend on the choice of local coordinate system for a point. It is essential that not every degenerate point is a bifurcation, while all critical points of the Morse function f(x) are bifurcations and are not topologically regular [1]. A point $x_0 \in M$ is called topologically regular if it is characterized by the fact that for a function f(x) on M there is an open neighborhood $U = U(x_0)$, which is homeomorphic to the direct product of the level surface f(x) by the surface given by $\{f^{-1}(a) \times [-\varepsilon, \varepsilon]\}$, where $a = f(x_0)$. If one considers one-dimensional spaces (curves, whose tangent space has dimension one), using the concept of simple point, whose definition requires just one local parameter and one generatrix for the ideal of its local ring.

Upon Morse transformation of the function f, the image of such representation in a group of linear transformations $E^\kappa$ is called the monodromy group of a singularity generated by reflections, namely, a loop (a closed geodesic line) around a critical, value acts as a reflection. Going around critical points of multiplicity m corresponds to the product of reflections in m mirrors, and for retained singularities – to going around the unique multiple value – a corresponding Coxeter element.

In a complex case, traversal of a critical value may be replaced by going around it, because a complex analog of an edge is a branched cover. Such a traversal corresponds to a non-singular manifold of level f=c, and monodromies, by definition, are various maps (not necessarily identical) of the manifold on itself. For example, the surface of a non-singular level $x^2 + y^2 = c$ ($c \neq 0$) in real case is determined by a circle, while topologically such surface is a cylinder in $E^4$. Going full circle around a critical value, the manifold, but not its separate points, return to the level [1].

If the number of critical values is fixed, then natural stratification of the base of miniversal deformation (small perturbations) of singularities is determined by expansion strata as a connected component of a set of parameters of such deformations. In what follows, the following features of such an approach are essential: the bifurcation diagram of zeros (deformation parameters for which the function has zero critical values – for simple singularities of the form $A_\kappa$, $D_k$, $E_6$, $E_7$, $E_8$ is diffeomorphic to variety of irregular orbits of reflection groups of the same name [23]. The $\mathbf{C}^n$ space allows locally for m-value covers branching along $\Sigma_2$ (m= (n+1)!, namely, for n=3: m=24 for $B_4$, m=2 for $C_4$ and m=6 for $F_4$ with symmetry of the cover space given by $A_2$. Thus, the choice of orbits of the group $F_4$ (with simultaneously appearing root lattices $D_4$ and $G_2$ by quasi-expansion) to construct polytopes is determined by the fact that upon projection (both smooth and non-smooth) of the front $F_4$, whereby one of its components is a cylinder over a hyper-surface in the base, locally ensures existence of a universal normal form [23].